# Autocrine signaling and quorum sensing: Extreme ends of a common spectrum

Berkalp A. Doğaner[1,2], Lawrence K. Q. Yan[1,2], and Hyun Youk[1,2,3]

[1]Department of Bionanoscience,
[2]Kavli Institute of Nanoscience, Delft University of Technology, Delft 2628CJ, the Netherlands
[3]Correspondence to: h.youk@tudelft.nl

- Cells often secrete and sense a signaling molecule to "talk" to each other. Autocrine signaling is one of the main forms of such communication. "Autocrine cell" refers to a cell that secretes a signaling molecule and makes its cognate receptor.
- Recent studies have shown that an autocrine cell can communicate with itself (self-communication) and communicate with other cells (neighbor-communication).
- Quorum sensing involves autocrine cells determining their population density due to the cells engaging in neighbor-communication without self-communication.
- A ubiquitous genetic circuit, called "secrete-and-sense circuit", controls the autocrine cell's ability to achieve self-communication, neighbor-communication (including quorum sensing), and a mixture of the two.
- Autocrine signaling and quorum sensing are two of many signaling modes enabled by the secrete-and-sense circuit.

**ABSTRACT**

"Secrete-and-sense cells" can communicate by secreting a signaling molecule while also producing a receptor that detects the molecule. The cell can potentially "talk" to itself ("self-communication") or talk to neighboring cells with the same receptor ("neighbor-communication"). The predominant forms of secrete-and-sense cells are self-communicating "autocrine cells" that are largely found in animals, and neighbor-communicating "quorum sensing cells" that are mostly associated with bacteria. While assumed to function independent of one another, recent studies have discovered quorum sensing organs and autocrine signaling microbes. Moreover, similar types of genetic circuits control many autocrine and quorum sensing cells. We outline these recent findings and explain how autocrine and quorum sensing are two sides of a many-sided "dice" created by the versatile secrete-and-sense cell.

**Secreting signaling molecules: A fundamental mode of communication**

Cells can communicate with each other by secreting signaling molecules that diffuse between them. Cells use a variety of receptors to detect the type and concentration of each extracellular signaling molecule. When the receptors bind to their cognate signaling molecules, they trigger cascades of intracellular signaling events that regulate diverse processes such as the growth and death of cells [1-3], differentiation [4-8], and gene expression [9-20]. We usually categorize cells that secrete signaling molecules into two types: those that engage in "autocrine signaling" and those that engage in "paracrine signaling" (Figure 1). In autocrine signaling, a cell secretes a signaling molecule and simultaneously makes a receptor for that molecule. Paracrine signaling involves two types of cells. One type of cell secretes a molecule without making a receptor for it and the other type of cell makes a receptor for the molecule without secreting the molecule. Along with contact-mediated signaling called "juxtacrine signaling" [21], autocrine and paracrine signaling are responsible for almost all known cell-cell communications in multicellular systems [22]. These modes of signaling have primarily been studied in mammalian systems. But recently, much progress has been made on studying paracrine signaling in populations of microbial cells such as bacterial biofilms and then extracting quantitative principles that apply to both mammalian systems (e.g., tissues) and microbial systems (e.g., biofilms) [23]. However many studies of autocrine signaling still mainly focus on mammalian systems and they typically exclude discussions on microbial cells, notably on how autocrine signaling may be related to quorum sensing. Quorum sensing, which allows the cells to "measure" their population density in order to make collective decisions, is one of the most well-known and ubiquitous forms of microbial communication.

While autocrine signaling and quorum sensing both involve cells that secrete a signaling molecule and express its cognate receptors, they have long been thought to be two disparate forms of signaling, likely because the two have seemingly different functions and purposes.

Autocrine signaling has been historically understood, albeit only recently demonstrated in live cells [24], to enable a single cell to "talk" to itself [25] whereas quorum sensing is designed for multiple cells to talk to each other but not for each cell to talk to itself [26]. In this sense, quorum sensing is similar to paracrine signaling in terms of its function because paracrine signaling is designed for a cell to talk to other cells but not to itself. On the other hand, quorum sensing is more similar to autocrine signaling than paracrine signaling in terms of its molecular parts (i.e., the same cell produces the receptor and the signaling molecule). Given these observations, it is natural to ask how autocrine signaling and quorum sensing might be related to each other both functionally and through evolution. Recently researchers have begun to concretely connect the two in terms of their common functions and features of the genetic circuits that control them [24]. Indeed, quorum sensing in mammalian organs [27] and autocrine signaling in microbes have been discovered [24], while additional work has shown that autocrine signaling and quorum sensing are two ends of a continuous spectrum of signaling modes that is spanned by a generic "secrete-and-sense cell" - a cell that secretes a signaling molecule and simultaneously makes its cognate receptor, but can talk to itself (like autocrine signaling), and talk to its neighbors (like quorum sensing and paracrine signaling) [24, 28]. These recent findings are causing a dismantling of the historically established barrier between researchers who have mainly studied quorum sensing in microbes (e.g., bacteria, yeasts) and researchers who have investigated autocrine signaling in metazoan cells (e.g., tumors, T-cells, embryos) [24]. Below, we review how researchers have traditionally thought about autocrine signaling and quorum sensing and describe recent studies that connect the two.

**Autocrine signaling: A cell that talks to itself**

One of the first descriptions of autocrine signaling arose when researchers proposed how tumor cells could originate in epithelial tissues in the 1980s [25]. It was known that many types of cells in healthy tissues secreted signaling molecules called Epidermal Growth Factors (EGFs) to regulate their proliferation (Figure 2A). It was hypothesized and later confirmed that when this autocrine signaling, which causes each healthy cell to stimulate its own growth by sensing its own growth factor molecule, is mis-regulated and thus over stimulates the cells, cells can grow uncontrollably and initiate tumors. Since then, researchers have found many examples of autocrine signaling in various mammalian cells [2, 29-34] (Figure 2B-D). In the human immune system, the naive T-helper cells use the molecules Interleukin (IL)-4 and interferon-$\gamma$ for autocrine signaling, in order to differentiate into one of two cell states (Th1 or Th2 cells) [35, 36]. The $CD4^+$ T-cells use autocrine signaling through IL-2 to control their proliferation and apoptosis [2] (Figure 2B). In early mammalian embryos including human embryos, a decreased level of autocrine signaling through the Platelet Activating Factor (PAF) ligand decreases the embryo's chances of survival (Figure 2C) [31-33]. Faulty regulations of autocrine signaling turn healthy cells into cancer cells and initiate the growth of tumors [29]. For example, renegade autocrine signaling through IL-6 can trigger lung adenocarcinoma in mice and humans. The somatic mutations in epidermal growth factor receptor (EGFR) in initially healthy mammary and lung cells cause the cells to secrete IL-6 at an abnormally high rate. The IL-6 then bind to the EGFRs on these cells, which then leads to highly activated STAT3 signaling in them. This in turn causes the cells to divide at an abnormally high rate, which leads to tumors [29] (Figure 2D).

Based on these examples, the prevailing belief has been that the primary purpose of autocrine signaling is for a cell to use its receptors to capture the signaling molecule that it had secreted so that it can "talk" to itself (which we call "self-communication") instead of sending the molecule to its neighboring cells to communicate with them (which we call "neighbor-communication"). The reasoning behind this is that for autocrine signaling, cells typically express a high abundance of receptors that have a high binding affinity for the signaling molecule [37]. Thus such a cell would have a high probability of capturing the molecules that it had just secreted. Since the molecule returns to the cell after being secreted, the cell would not be able to communicate with its neighboring cells. According to this scenario, autocrine signaling would require only a single cell and such a cell would be able to rapidly respond to its own signaling molecule [38-40]. Moreover this scenario could provide a plausible reason for why autocrine signaling is ubiquitous in controlling proliferation of cells through growth factors and in embryogenesis. In both embryogenesis and signaling through growth factors, cells have to rapidly undergo changes in their growth, gene expression, and differentiation. A cell that can quickly capture its own signaling molecule would be able to more rapidly respond to the molecule than a cell that relied on a molecule from other cells (i.e., a receiver cell in paracrine signaling) because the molecule would need a shorter distance to travel than in paracrine signaling. However, this idea of pure self-communication has persistently posed two questions that are only now being resolved: i. why would a cell go through the many steps of producing and secreting a molecule if it only wanted to communicate with itself. ii. Why does the cell not rely entirely on intracellular signals? The other challenge has been that for many years, it was difficult to experimentally show that autocrine signaling involved self-communication in individual cells. It was not until recently that researchers performed measurements of gene expression with single cell resolution to definitively prove that autocrine signaling enables pure self-communication without any neighbor-communication [24]. These measurements and nascent theoretical studies [41-43] are now beginning to resolve the aforementioned two questions. We will elaborate these resolutions after first reviewing some basics of quorum sensing.

**Quorum sensing: A cell that talks to other cells**

Quorum sensing is a form of signaling in which a cell secretes a signaling molecule in order to communicate with other cells (i.e., engaging in a pure "neighbor-communication") in a way that depends on the density of the cell population. It has been the main paradigm for understanding multicellular behaviors and communication among bacteria and microbial eukaryotes [26]. Quorum sensing triggers coordinated and collective actions such as all cells in the population turning on the same gene once the cell population density is above a certain threshold value. In this way, one can consider quorum sensing to be paracrine signaling that is activated by cells when the density of the cell population is above a certain threshold, whereas the cells do not communicate with self or neighbors when the population density is below the threshold. Quorum sensing is usually used when the benefits of cooperative actions outweigh the benefits of each cell acting autonomously [26].

Microbial cells, rather than being just autonomous individuals, use quorum sensing to accomplish tasks as a collective entity. For example, the marine bacteria *Vibrio fischeri* reside inside the "light organ" of the Hawaiian Bobtail Squid *Euprymna scolopes* and uses quorum sensing through the secreted molecule AHL (Acyl Homoserine Lactone) to produce light inside the squid (a phenomenon called bioluminescence) (Figure 3A) [44-47]. When the population density of the *Vibrio fischeri* cells is low, the concentration of the secreted AHLs remains low. As the population density increases, so does the concentration of AHL. At a certain population density, the concentration of AHL goes above a genetically encoded threshold, which then turns on downstream genes that lead to bioluminescence [26]. Researchers have also engineered genetic circuits in *Escherichia coli* cells so that the cells can quorum sense through AHL, which has provided investigation into population-level behaviors, such as population density control of *E. coli*'s rate of death [48, 49].

Over the last decade, numerous studies have combined mathematical modeling with fluorescence based methods such as time-lapse microscopy for measuring gene expression in single cells to reveal how individual cells encode the threshold for quorum sensing and the kinds of genetic circuits that can achieve quorum sensing [50-53]. In the soil amoeba *Dictyostelium discoideum*, a special form of quorum sensing, called "dynamic quorum sensing", causes the cells to transition from unicellular individuals to macroscopic aggregates called "fruiting bodies" (Figure 3B) [54-57]. For example, amoeba cells continuously secrete Pre-Starvation Factor (PSF). The concentration of PSF increases as the density of starving cells increases [57]. When the PSF concentration reaches a certain threshold, the cells respond by turning on the expression of a set of genes that eventually trigger secretion of the chemo-attractant, cyclic Adenosine Mono-Phosphate (cAMP). cAMP secretion is dynamically regulated by the density of the amoeba, with a positive feedback that regulates the secretion of cAMP (i.e., cells increase their average secretion rate of cAMP as they sense more cAMP). This regulation eventually leads to the aggregation of cells into fruiting bodies [7, 57, 58]. This example suggests that quorum sensing may have been crucial in the evolution of multicellularity.

Quorum sensing is also seen in cooperative and commensal relationships among different species [59-63]. Moreover researchers are currently investigating inhibitors that disrupt quorum sensing in bacteria (i.e., "quorum quenching") as an alternative to antibiotics to which many bacteria have developed resistance [60-65], to treat cancer [60, 66], and to treat wounds [67].

Based on these examples, quorum sensing can be considered a form of paracrine signaling that depends on the density of the cell population despite that quorum sensing cells produce both a signaling molecule and its receptor, which is more similar to autocrine cells. Typically, receptors for the signaling molecule used in quorum sensing have a low binding affinity for the molecule [26]. Moreover, these receptors tend to be expressed in low abundance. Thus microbes that quorum sense would only be able to detect the presence of the signaling molecule when there is a sufficiently high density of it, which would occur only when the cell population density is sufficiently high [26].

**A secrete-and-sense cell uses one molecule to talk to itself and to other cells**

One of the main goals of systems biology is to connect seemingly disparate biological systems under common quantitative principles. Recent studies of quorum sensing and autocrine signaling are pointing towards such unification between the two modes. For one, researchers are now discovering quorum sensing in metazoan cells and in animal populations (Figure 3C-D). For example, researchers have recently found that hair follicles underneath the mouse skin regenerate damaged hairs only if the density of damaged hairs is above a certain threshold (i.e., number of plucked hairs per unit area of skin) (Figure 3C) [27]. This constitutes an example of quorum sensing at the level of a whole organ (hair follicle). Researchers have also found that the ants, *Temnthorax albipennis*, use quorum sensing to migrate to their nest. Namely, each ant counts the rate at which it encounters other ants as the ants roam about in search of a new nest. Once each ant's rate of encounter goes above a certain threshold, the ants collectively migrate to the region where each of them (and thus everyone) experiences an encounter rate that is above the threshold. This rate informs their new nesting site (Figure 3D) [68]. Similar group decisions have been observed in honeybees migrating to their nests [69].

Adding towards the trend of unification has been the recent demonstration of autocrine signaling in microbial cells [24]. A recent study has demonstrated autocrine signaling in engineered budding yeasts and in the process, proposed a fundamental connection between autocrine signaling and quorum sensing [24]. The study engineered a simple genetic circuit in budding yeast cells that caused the cells to secrete a mating pheromone ($\alpha$-factor) and express a receptor (Ste2) for that pheromone [70]. The different concentrations of the mating pheromone caused the cells to express different amounts of a fluorescent protein (but not mate with each other). Four parts of the "secrete-and-sense" genetic circuit could also be tuned independently of each other over a wide range and demonstrated various "social behaviors" that the engineered yeast cells could achieve (Figure 4). These factors were: i. The expression level of the receptor, ii. Secretion rate of the $\alpha$-factor, iii. The expression level of a protease that actively degraded the $\alpha$-factor outside the cell, and iv. The strength of a positive feedback that caused the cells to increase the secretion rate of the $\alpha$-factor as the sensed concentration of the $\alpha$-factor increased. Through this tuning, the study showed that autocrine signaling and quorum sensing are merely two extreme ends of a continuous spectrum of a signaling modality and proposed that a more generic "secrete-and-sense cell" could span this "sociability" spectrum (Figure 4). Namely, autocrine signaling is the "asocial" end of the sociability spectrum while quorum sensing is the "social" end of the sociability spectrum (Figure 4). The generic secrete-and-sense cell, which is any cell (mammalian or bacterial) whose genetic circuit contains the four parts mentioned above would span the rest of this spectrum by tuning these four parts of the secrete-and-sense circuit [24].

In other recent studies [28, 42], researchers have developed mathematical models to reveal how the parameters of secrete-and-sense circuits could be tuned to allow the cells to communicate with each other, thus achieving efficient neighbor-communication that is the basis of quorum sensing. In a recent study [42], researchers developed a mathematical model to analyze the secrete-and-sense circuit in the T-cells of our immune system. Their model showed that a group of secrete-and-sense T-cells can compete for the same global pool of the secreted IL-2. Their model predicted this by tuning the "activation threshold", which is the concentration of the IL-2 at which the cells can switch on their genes that affect their proliferation rate. Thus the researchers showed that in a population of polyclonal T-cells with distinct activation thresholds, the cells could either compete or cooperate with each other. Recent experiments [2, 24] have confirmed key predictions of this model and affirmed that secrete-and-sense cells could achieve density dependent paracrine signaling (i.e., quorum sensing).

These and other recent studies use experiments [2, 24] and mathematical models [28, 42] to show that the secrete-and-sense circuit motif (Figure 4) allows cells to realize autocrine signaling and density-dependent paracrine signaling (i.e., quorum sensing) at different times to achieve distinct goals. They demonstrate that there is no rigid boundary between quorum sensing and autocrine signaling. This division is an historical artifact of researchers in different disciplines having studied quorum sensing and autocrine signaling as separate phenomena. By considering both quorum sensing and autocrine signaling at the same time, recent studies have shown that there is a continuous spectrum between the two modes of signaling [2, 24, 28, 42]. A mammalian or microbial cell can use autocrine signaling and quorum sensing simultaneously. Cells can tune their degree of autocrine signaling and their degree of quorum sensing by tuning each of the four elements of its genetic circuit to regulate respectively the amount of self-communication and neighbor-communication (Figure 4). The cell can continuously tune these two degrees of communication to realize a "spectrum" of signaling modes (i.e., a self-communication and neighbor-communication). A fundamental trade-off in this spectrum however, is that a high degree of neighbor-communication comes at the cost of lowering the degree of self-communication and vice versa [24].

**Concluding Remarks**

We have outlined nascent studies that demonstrated autocrine signaling to be more closely related to quorum sensing than was previously thought. We can consider quorum sensing to arise from the ability of cells that are conventionally called "autocrine cells" to engage in pure neighbor-communication (Figure 4) as in paracrine signaling. Conversely, cells that quorum sense can engage in self-signaling, as in autocrine signaling, by tuning their secretion and sensing of the signaling molecule. Importantly, both autocrine signaling cells and quorum sensing cells can tune their genetic circuits to realize a mixture of self-communication and neighbor-communication (Figure 4). This unified view of autocrine signaling and quorum sensing shows the deficiency of the term "autocrine signaling" as it is conventionally used in literature and textbooks [71, 72], which refers to cells that secrete a signaling molecule and express its cognate receptor but does not distinguish between whether such a cell communicates with itself and/or with its neighbors. Cells that have been traditionally called "autocrine cell" can engage in paracrine signaling, including its density dependent form that we call quorum sensing [24, 28]. But in light of the recent findings outlined in this review, we should focus on which cell communicates with which other cell when referring to metazoan and bacterial autocrine cells.

Despite these recent progresses, many important questions still remain (see Outstanding Questions). An outstanding question is the evolutionary origin of autocrine signaling. One possible answer is that quorum sensing first originated in bacteria, and then the early forms of multicellular organisms (i.e., animals) inherited the secrete-and-sense circuit of the bacteria that was tuned for quorum sensing. The animals could have then tuned each of the four main elements of the secrete-and-sense circuit (e.g., through mutations in the regulatory regions) to achieve pure autocrine signaling or a mixture of self- and neighbor-communication. Future work that investigates this possibility will likely yield a more comprehensive view of secrete-and-sense cells.

It is also interesting to investigate the possibility that microbial and metazoan cells might tune their levels of autocrine signaling and quorum sensing at different times to suit different needs. For example, at the beginning of the development of an embryo or a tissue, autocrine signaling may be used in order to achieve a great biomass and population density within a short time period, and then switch to quorum sensing to enable the cells within the embryo or the tissue to coordinate their behaviors. The different time scales involved in the effects of autocrine signaling and quorum sensing may be worth investigating in the future.

Finally, many mammalian tissues, for example the islets of Langerhans of the pancreas, consist of defined spatial arrangements of distinct cell types that engage in both autocrine and paracrine signaling (e.g., the beta-cells of pancreas use insulin for autocrine signaling) [73]. Therefore, an interesting question is how the physical structures of various tissues affect the degree of autocrine and potentially of quorum sensing in the tissues.

We envision that the above types of studies will reveal deep quantitative principles that govern secrete-and-sense cells and cellular communication in general.

**ACKNOWLEDGMENTS**

We thank D. Gomez-Alvarez, E. Helguero and A. Ravensbergen for insightful suggestions. H.Y. is supported by a NWO NanoFront Grant. We apologize to researchers whose work we have missed in citing.

## Two fundamental modes of cellular communication

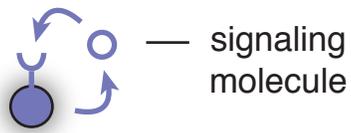

autocrine signaling — signaling molecule

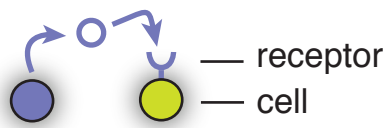

paracrine signaling — receptor — cell

Fig. 1

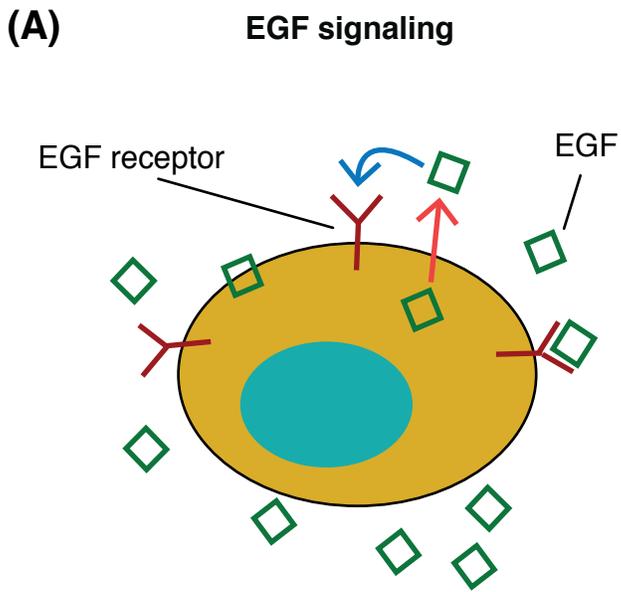
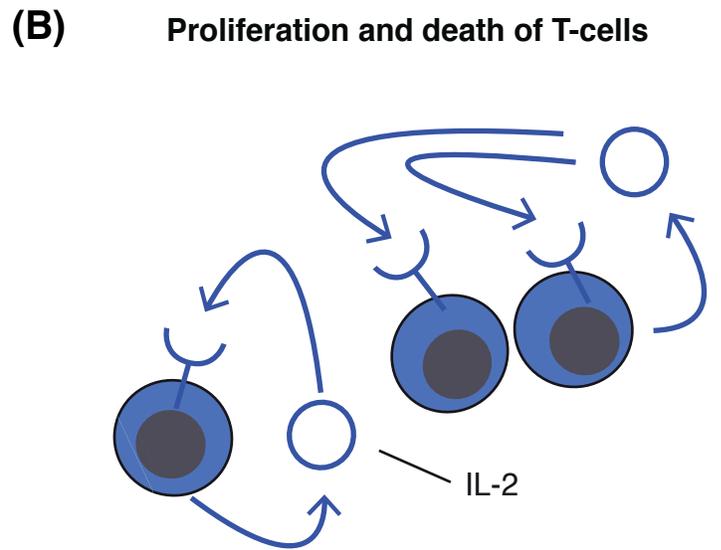
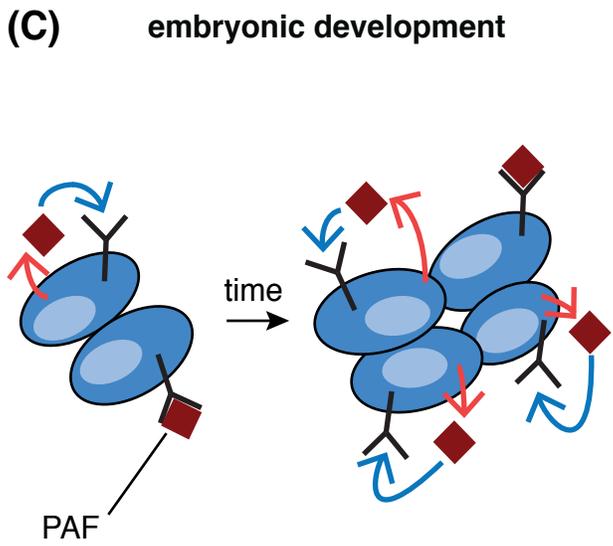
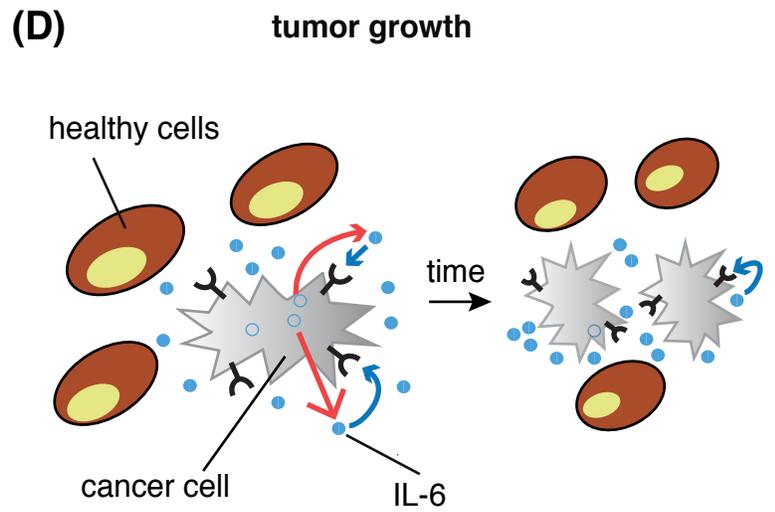

Fig. 2

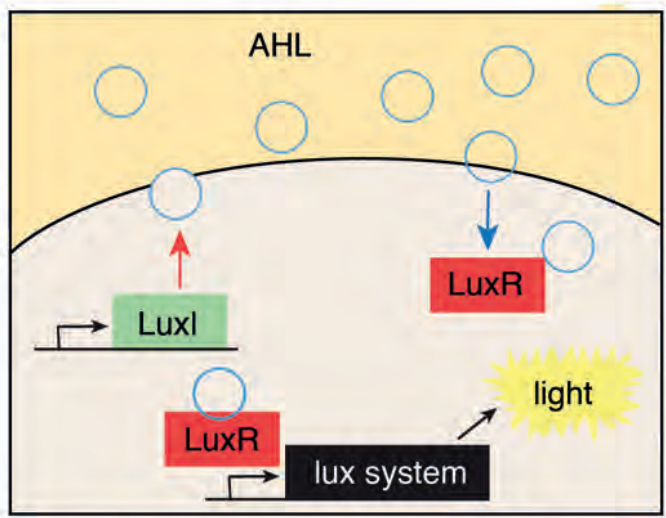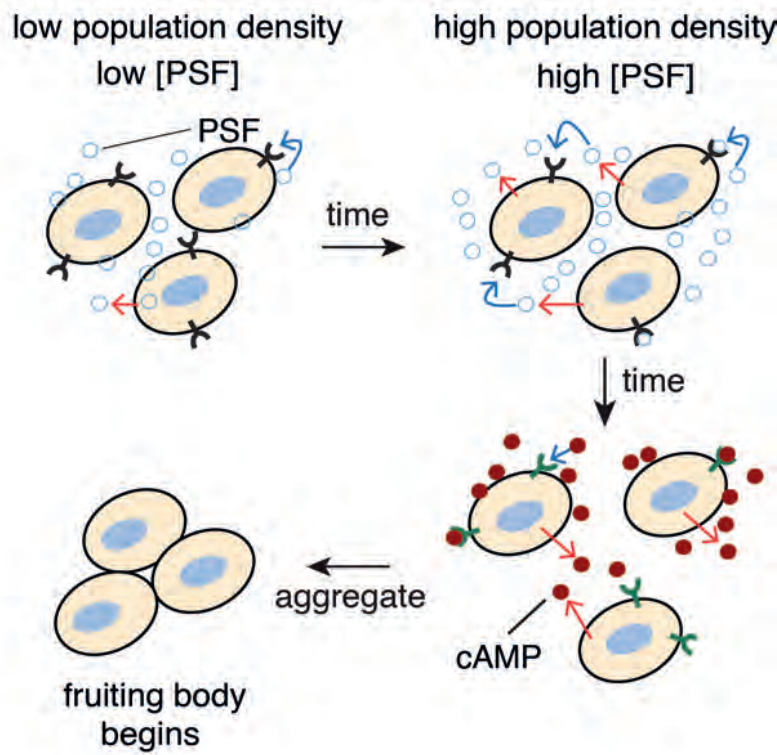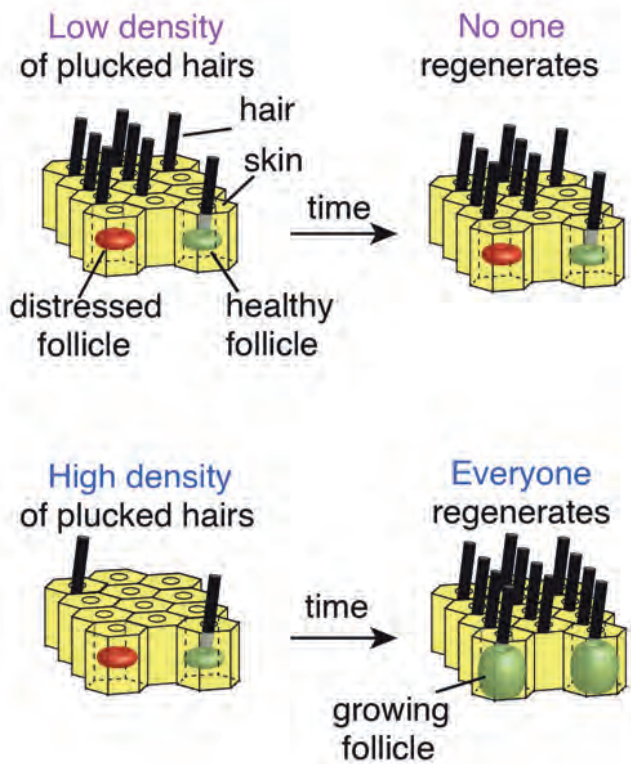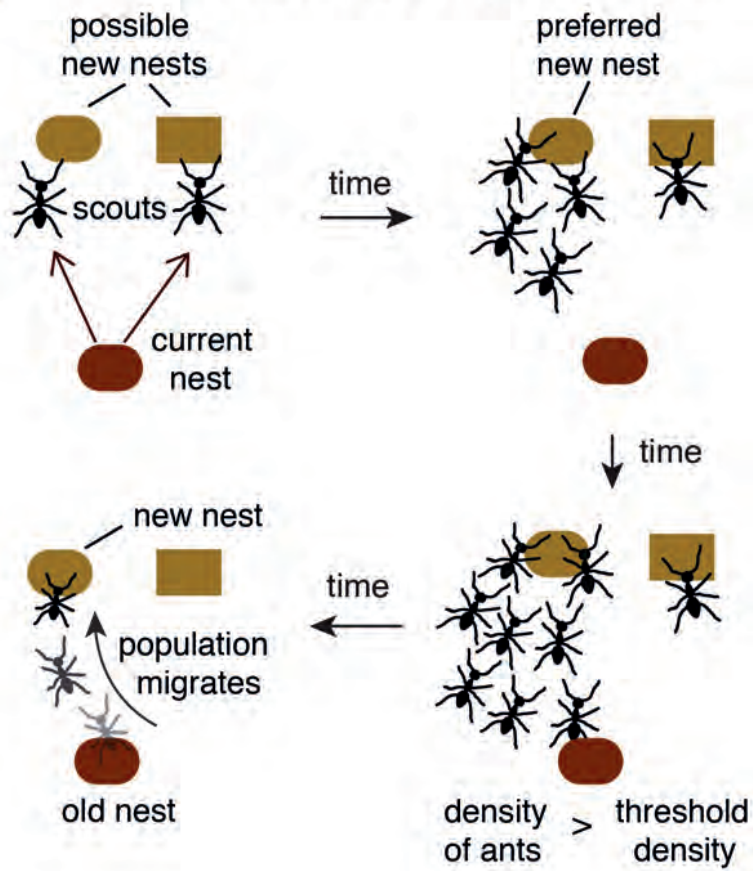

Fig. 3

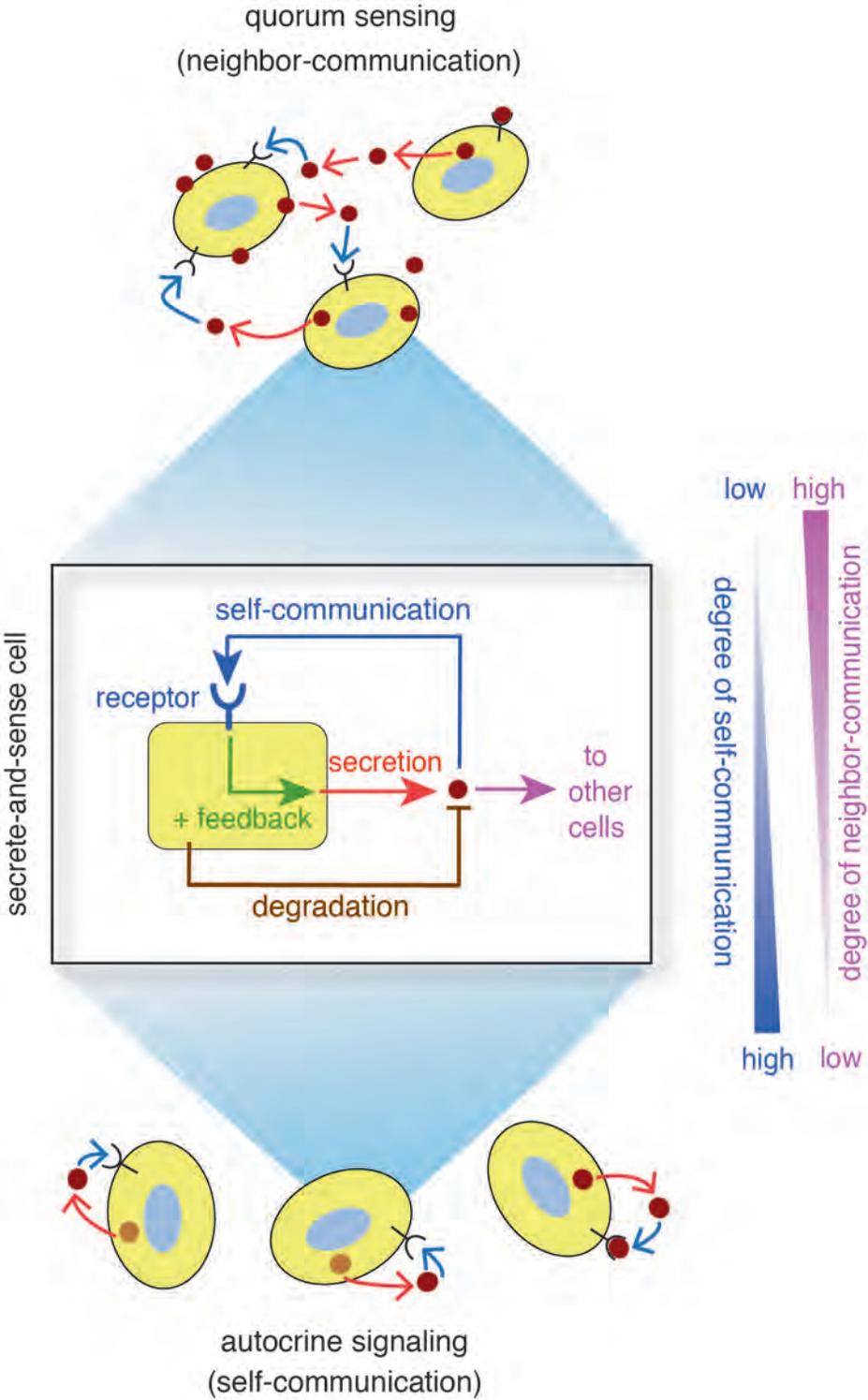

Fig. 4

**Figure captions:**

**Figure 1. Autocrine and paracrine signaling as two fundamental means of cellular communication.** Cells often communicate by secreting a signaling molecule. Autocrine signaling and paracrine signaling are the two fundamental and ubiquitous modes of communication through a secreted signaling molecule. In autocrine signaling, a cell secretes a signaling molecule and simultaneously makes a receptor that captures the molecule. In paracrine signaling, one type of cell secretes a signaling molecule without making its cognate receptor while another type of cell makes the cognate receptor without secreting the molecule.

**Figure 2. Examples of autocrine signaling. (A)** Epidermal growth factor (EGF) and its receptor (EGFR) in epithelial cells. **(B)** The $CD4^+$ T cells use autocrine signaling through IL-2 to control their proliferation and apoptosis. **(C)** Autocrine signaling, for example through the Platelet Activating Factor (PAF), plays an important role in the development of mammalian embryos. A decreased amount of the PAF ligand reduces the embryo's chances of surviving through development. **(D)** Faulty autocrine signaling through IL-6 causes uncontrolled growth in healthy cells and initiates tumor formation in the early stages of breast and lung cancers.

**Figure 3. Examples of quorum sensing. (A)** When the population density of the marine bacteria *Vibrio fischeri* is small, the concentration of the secreted AHL (Acyl Homoserine Lactones) is low. An increase in the cell population density causes the extracellular concentration of AHL to rise. When the concentration of AHL goes above a certain threshold, the cells generate light through the "Lux system". **(B)** The soil amoeba *D. discoideum* continuously secrete Pre-starvation factor (PSF). The concentration of PSF increases as the density of starving cells rises. When the PSF concentration reaches a certain threshold, the amoeba respond by secreting cyclic Adenosine Mono-Phosphate (cAMP). This eventually leads

to the cells aggregating into fruiting bodies. **(C)** Hair follicles regenerate damaged hairs if and only if the density of damaged hairs is above a certain threshold. **(D)** The ant (*Temnthorax albipennis*) counts the rate at which it encounters other ants as it walks around in search of a new nesting site. Once the rate at which each ant encounters other ants goes above a certain threshold, the ants collectively migrate to the region in which the encounter rate is above the threshold to establish their new nest.

**Figure 4. Autocrine signaling and quorum sensing are two extreme ends of a spectrum of signaling modalities that secrete-and-sense cells can realize.** A secrete-and-sense cell can tune each of the four main elements of its genetic circuit to control its degree of self-communication (i.e., extent of autocrine signaling) and its degree of neighbor-communication (i.e. extent of quorum sensing). The cell can continuously tune these two degrees of communication to realize a "spectrum" of signaling modes (i.e., a mixture of autocrine signaling and quorum sensing). A fundamental trade-off is inherent in this spectrum: A high degree of neighbor-communication comes at the cost of lowering the degree of self-communication and vice versa.